\documentclass[10pt]{article}
\usepackage{a4}
\usepackage{amssymb}
\usepackage[dvips]{epsfig}
\usepackage[dvips]{graphics}
\begin{document}

\title{Bose--Einstein Condensation and Free DKP field}
\author{R. Casana, V. Ya. Fainberg\footnote{
Permanent address: P. N. Lebedev Institute of Physics, Moscow,
Russia }\,\,, B.M. Pimentel and J. S. Valverde \vspace{.2cm}\\
{\small Instituto de F\'{\i}sica Te\'orica, Universidade Estadual
 Paulista} \vspace{-.1cm}\\
{\small Rua Pamplona 145, CEP 01405-900, S\~ao Paulo, SP, Brazil}}
\date{}
\maketitle

\begin{abstract}
The thermodynamical partition function of the Duffin--Kemmer--Petiau
theory is evaluated using the imaginary-time formalism of quantum
field theory at finite temperature and path integral methods. The
DKP partition function displays two features: (i) full equivalence
with the partition function for charged scalar particles  and
charged massive spin 1 particles; and (ii) the zero mode sector which
is essential to reproduce the well-known relativistic Bose--Einstein
condensation for both theories.
\end{abstract}

\vskip 0.7cm
{\small \hspace{-0.625cm} Dedicated to Professor S\'\i lvio
Roberto de Azevedo  Salinas on the  occasion of his 60th birthday.}

\section{Introduction}

Bose--Einstein condensation \cite{bose,einstein} (BEC) is a
phenomenon long associated to the liquid helium studies ($^4$He and
$^3$He--$^4$He mixtures). However, recent research deals with
BEC shown that it  touches several areas of modern physics
\cite{bec1x}: in thermodynamics \cite{huang} BEC occurs as a phase
transition from gas to a new state of matter, quantum mechanics
view BEC as a matter-wave coherence \cite{bec1} arising from the
overlapping de Broglie waves of the atoms and draw an analogy between
conventional and ``atom lasers" \cite{bec2}, quantum statistics
explain BEC as more than one atom sharing a phase space cell, in
the quantum theory of atomic traps \cite{bec3} many atoms condense
to the ground state of the trap, in quantum field theory BEC is
commonly related to spontaneous symmetry breaking
\cite{kapusta1,weldon,bernstein}.

In quantum field theory BEC the complex scalar field was and is
used for studying the thermodynamical properties of physical
systems composite of bosonic particles with spin 0. However, at
zero temperature, there is an alternative way to study not only
the properties for complex scalar field but also the complex
vectorial field that it is known as the massive
Duffin--Kemmer--Petiau (DKP) theory \cite{dkp}.

At zero temperature, one important question concerning DKP theory
is about the equivalence or not between its spin 0 and 1 sectors
and the theories based on the second order Klein-Gordon (KG) and
Proca equations, respectively. From the beginning of the 50's the
belief on this equivalence was perhaps the principal reason for
the abandon of DKP equation in favor of KGF and Proca ones.
However, in the 70's this supposed equivalence began to be
investigated in several  situations involving breaking of symmetries
and hadronic processes, showing that in some cases DKP and KG
theories can give different results (for a historical review of
the development of DKP theory until the decade of 70's see the
reference \cite{Nieto}). Moreover, DKP theory follows to be richer
than the KG one with respect to the introduction of interactions.
In this context, alternative DKP-based models were proposed for the
study of meson-nucleus interactions, yielding a better adjustment to
the experimental data when compared  to the KG-based theory
\cite{Kalbermann}. In the same direction, and guided for the formal
analogy with Dirac equation, approximation techniques formerly
developed in the context of nucleon-nucleus scattering were
generalized, giving a good description for experimental data of
meson-nucleus scattering. The  deuteron-nucleus scattering was also
studied using DKP, motivated by the fact that this theory suggests
a spin 1 structure from combining two spin $\frac{1}{2}$
\cite{Clark,Kozack}.

Recently there have been a renewed interest in DKP theory. For
instance, it has been studied in the context of QCD \cite{Gribov},
covariant hamiltonian dynamics \cite{Kanatchikov}, in the causal
approach \cite{Lunardi 3}, in the context of
five-dimensional galilean covariance \cite{Esdras}, in the scattering
$K^+$-nucleus \cite{Prog}, in curved space-times
\cite{Lunardi 1,fainberg0,Lunardi 4,trabs=1}, etc.

These examples, among others in literature, in some cases break the
equivalence between the theories based on DKP and KG/Proca equations,
such as in \cite{Kalbermann} or in Riemann-Cartan space--times
\cite{Lunardi 4,trabs=1}. Nevertheless, the question about the
equivalence or not still lacks a complete answer nowadays.

As above mentioned, all accomplished studies on the DKP theory were
made at zero temperature. The aim of this work is to study the
thermodynamics and Bose--Einstein condensation for the massive
charged particles by using the DKP theory. In section 2 we  present
the massive DKP theory in the Minkowski space-time and we make a
r\'esum\'e of the constraint analysis of the model which can be seen
in \cite{pimentel-1}, where it has been shown that all the photon
Green's functions coincide in DKP and KGF theories. In the section 3
we introduce the partition  function  and the generating functional
of the correlation functions of the theory, and we specialize,
separately, the spin 0 and spin 1 sector for the explicit calculation
of the respective partition functions. We also analyze explicitly the
zero mode contribution in both cases. In the section 4 we gives our
conclusions and perspectives.

\section{The Duffin-Kemmer-Petiau theory}

The Duffin-Kemmer-Petiau equation \cite{dkp} in Minkowski
space--time is given by
\begin{eqnarray} \label{eq1x}
i\beta^{\mu}\partial_{\mu}\,\psi -m\psi =0,
\end{eqnarray}
where the matrices $\beta ^{\mu}$ obey the DKP algebra,
\begin{eqnarray} \label{eq2x}
\beta^{\alpha}\beta^{\mu}\beta^{\nu}+\beta^{\nu}\beta^{\mu}
\beta^{\alpha}=\beta^{\alpha}\eta ^{\mu\nu}+\beta^{\nu}
\eta^{\mu\alpha}
\end{eqnarray}
with $\eta^{\mu\nu}$ being the metric tensor of Minkowski space-time
with signature $\left( +---\right) $. The $\beta^\mu$ are singular
matrices which have only three irreducible representations of
dimensions 1, 5 and 10. The first one is trivial, having no physical
meaning and the other two correspond to fields of spin 0 and 1,
respectively.

The DKP equation given in (\ref{eq1x}) is obtained from the
following Lagrangian density
\begin{eqnarray} \label{eq3x}
{\cal L}\;=\;\frac{i}{2}\;\overline\psi\beta^\mu\partial_\mu \psi
-\frac{i}{2}\;\partial_\mu\overline\psi\beta^\mu \psi -
m\overline\psi \psi\,,
\end{eqnarray}
where $\overline\psi=\psi^\dag\eta^0$ and $\eta^0=2(\beta^0)^2-1$,
with $(\eta^0)^2=1$

In \cite{pimentel-1}, the Hamiltonian canonical approach to the
DKP theory is developed in the components and the matrix forms. We
make a r\'esum\'e of the matrix form of this analysis to follow. The
canonical conjugate momenta are
\begin{eqnarray}\label{eq4x}
p_\psi =\frac{i}{2}\;\psi^\dag\beta^0 \quad,\quad p_{\overline\psi}
= -\frac{i}{2}\;\beta^0 \psi \;.
\end{eqnarray}

The Hamiltonian density is given by
\begin{eqnarray}\label{eq5x}
\mathcal{H} &  \!\!=&\!\!\! -\frac{i}{2}\;  \overline\psi
 \beta^k \partial_k \psi  +\frac{i}{2}\;
\partial_k\overline\psi\beta^k  \psi +
m\overline\psi\psi\;.
\end{eqnarray}

There is a set of second--class constraints
\begin{eqnarray}\label{eq6x}
\overline\theta=p_\psi-\frac{i}{2}\, \psi^\dag\beta^0 \quad,\quad
\theta=p_{\overline\psi}+\frac{i}{2}\,\beta^0 \psi\;,
\end{eqnarray}
\begin{eqnarray}\label{eq7x}
{\omega}=\mathbf{M}(i\,\beta^k\partial_k \psi-m\psi ) \quad,\quad
\overline\omega=\left(i\,\partial_k\overline\psi \beta^k+m
\overline\psi\right)\mathbf{M}\;,
\end{eqnarray}
where ${\bf M}=1-(\beta^0)^2$ is a singular matrix.


\section{The Partition function}

The massive DKP Lagrangian density (\ref{eq3x}) has a global $U(1)$
symmetry
\begin{eqnarray}
\psi&\rightarrow&\psi'=e^{i\alpha}\psi \nonumber\\
\overline\psi&\rightarrow&\overline\psi'=e^{-i\alpha}\overline\psi
\end{eqnarray}
whose conserved current is $j^\mu=\overline\psi\beta^\mu\psi$, and
the respective conserved charge
\begin{eqnarray}
Q=\int\!\!d^3{\bf x}\;j^0(x)\;.
\end{eqnarray}
Then the partition function in the imaginary-time formalism for the
massive DKP field is
\begin{eqnarray} \label{eq10x}
Z&\!\!\!\!=&\!\!\!\!N(\beta)\!\!\int\!{\cal D}p_\psi\, {\cal
D}p_{\overline\psi} \int_{_{\hspace{-0.5cm}\mbox{\footnotesize
periodic}}} \hspace{-0.2cm}{\cal D}\psi \,{\cal D}\overline\psi\;
\delta\left(\theta\right)\delta\left(\overline \theta \right)
\delta\left(\omega\right)\delta\left(\overline\omega\right)\\
& & \hspace{3cm}\exp\left\{\int_0^\beta\!\!d\tau\!\!\int\!d^3{\bf
x} \left[\frac{}{} i p_\psi\, \partial_\tau \psi+i\partial_\tau
\overline\psi\, p_{\overline\psi} -{\cal H}+\mu j^0\right]\right\}
\nonumber ,
\end{eqnarray}
where $N(\beta)$ is an infinite normalizing factor which will be
determined later \cite{bernard1}. We integrate over periodic DKP
field due to its bosonic character.

To compute the correlation functions of the theory we couple the
DKP field to external sources $\overline\eta$ and $\eta$, thus
the generating functional is read
\begin{eqnarray} \label{eq11x}
Z[\eta,\overline\eta]&\!\!\!\!=&\!\!\!\!N(\beta) \!\!\int\!{\cal
D}p_\psi\, {\cal D} p_{\overline\psi}
\int_{_{\hspace{-0.5cm}\mbox{\footnotesize periodic}}}
\hspace{-0.2cm}{\cal D}\psi \,{\cal D}\overline\psi\;
\delta\left(\theta\right)\delta\left(\overline \theta \right)
\delta\left(\omega\right)\delta\left(\overline\omega\right)\\
& & \hspace{0.5cm}\exp\left\{\int_0^\beta\!\!d\tau\!\!
\int\!d^3{\bf x} \left[\frac{}{} i p_\psi\, \partial_\tau
\psi+i\partial_\tau \overline\psi\, p_{\overline\psi} -{\cal
H}+\mu j^0+\overline\eta\psi+\overline\psi\eta \right]\right\}
\nonumber.
\end{eqnarray}

After the integration on the canonical conjugate momenta, we
define  the matrix $\beta^\tau\equiv i\beta^0$; and the constraints
$\omega$ and $\overline\omega$ can be rewritten as
\begin{eqnarray}
\omega = {\bf M\,D}\psi \quad,\quad \overline\omega = \overline{{\bf
D}\psi}\,{\bf M} \nonumber
\end{eqnarray}
due the property $\beta^0{\bf M}=0={\bf M}\beta^0$, and we have used
a short notation,
\begin{eqnarray}
\mathbf{D}\psi = i\beta^a\partial_a\psi-m\psi \quad,\quad
\overline{\mathbf{D}\psi} &\!\!\!=&\!\!\!i\,\partial_a\overline
\psi \beta^a+m\overline\psi\,,
\end{eqnarray}
where $a=\tau,1,2,3$. With all definitions noted above, the
generating functional (\ref{eq11x}) appears to be
\begin{eqnarray} \label{eq17x}
Z [\eta,\overline\eta]=N(\beta)
\int_{_{\hspace{-0.5cm}\mbox{\footnotesize periodic}}}
\hspace{-0.2cm}{\cal D}\psi \,{\cal D}\overline\psi\;
\delta\left(\mathbf{M}\,\mathbf{D}\psi\right)
\delta\left(\overline{\mathbf{D}\psi}\,\mathbf{M}\right) \; e^{S}
\end{eqnarray}
where the action $S$ is given by
\begin{eqnarray}\label{ft15}
S= \int_0^\beta\!\!d\tau\!\!\int\!d^3{\bf x}\left( \frac{1}{2}\;
\overline\psi \,\mathbf{D}\psi- \frac{1}{2}\;\overline{\mathbf{D}
\psi}\,\psi +\mu \overline\psi \beta^0\psi + \overline\eta\psi+
\overline\psi\eta\right).
\end{eqnarray}
After the exponentiation of the $\delta$--functions, we use the
periodicity condition  over the fields to make some integration by
parts, and get the following expression for the generating
functional
\begin{eqnarray} \label{ft16}
Z [\eta,\overline\eta]=N(\beta)\int_{_{\hspace{-0.5cm}
\mbox{\footnotesize periodic}}} \hspace{-0.2cm}{\cal D}\psi
\,{\cal D}\overline\psi\; \,{\cal D}\rho \,{\cal
D}\overline\sigma\; e^{S^1}
\end{eqnarray}
where the auxiliary fields $\rho$ and $\sigma$ were used to
exponentiate the constraints; and the action $S^{1}$ is given by
\begin{eqnarray}\label{ft17}
 S^1=\int_0^\beta\!\!d\tau\!\!\int\!d^3{\bf x} \left[\frac{}{}
\overline\psi \left(\mathbf{D}+\mu \beta^0\right)\psi
-i\,\overline{\mathbf{D M}\sigma}\,\psi -i\,\overline{\psi}
\,\mathbf{D}\mathbf{M} \rho+ \overline\eta\psi+
\overline\psi\eta\frac{}{}\right]
\end{eqnarray}

Into the generating functional (\ref{ft16}), we make the following
change of variables,
\begin{eqnarray}
\psi \rightarrow  \psi +\psi_{s} \quad,\quad \overline\psi
\rightarrow \overline\psi +\overline \psi_{s}
\end{eqnarray}
such the fields $\psi_{s}$ and $\overline\psi_{s}$ satisfy the
following equations
\begin{eqnarray}
\left(\mathbf{D}+\mu \beta^0\right)\!\psi_{s}-i\,
\mathbf{D}\mathbf{M}\rho+\eta\;=\;0 \quad;\quad
-\overline{\mathbf{D}{\psi_s}}+\mu \overline\psi_s\beta^0 -
i\,\overline{\mathbf{D M}\sigma}+\overline\eta \;=\;0\;.\nonumber
\end{eqnarray}
And after it, we carry out another change of variables on the
auxiliary fields $\rho$ and $\sigma$,
\begin{eqnarray}
\overline\sigma\rightarrow\overline\sigma+\overline\sigma_s\quad,
\quad \rho\rightarrow\rho+\rho_s
\end{eqnarray}
where the fields $\rho_s$ and $\sigma_s$ satisfy the equations
\begin{eqnarray}
m{\bf M}\rho_s-i{\bf M}\eta=0 \quad;\quad m\overline\sigma_s{\bf
M}+i\overline\eta{\bf M}=0  \nonumber\;.
\end{eqnarray}

Thus, after some additional calculations in the generating
functional in (\ref{ft16}), we get
\begin{eqnarray} \label{eq32x}
Z [\eta,\overline\eta]= \exp\left\{-\int_0^\beta\!\!\!
d\tau\!\!\int\!d^3{\bf x} \;\overline\eta \left(
\mathbf{D}+\mu\beta^0 \right)^{-1}\!\!\eta+\frac{1}{m}\;
\overline\eta{\bf M}\eta\right\}\;Z\;,
\end{eqnarray}
where $Z$ is the partition function given in (\ref{eq10x}) which,
after all operations made above, it is read as
\begin{eqnarray}
Z=N(\beta)\! \int_{_{\hspace{-0.5cm}\mbox{\footnotesize periodic}}}
\hspace{-0.4cm}{\cal D}\psi \,{\cal D}\overline\psi\; {\cal
D}\rho\,{\cal D}\overline\sigma\;\exp\left[\int_0^\beta \!\!\!
d\tau\!\!\int\!d^3{\bf x} \;\overline\psi({\bf D}+ \mu\beta^0)\psi
+m\,\overline\sigma{\bf M}\rho\,\right].
\end{eqnarray}

We can note that the functional integration over the fields
$\overline\sigma$ and $\rho$ give an infinity quantity, due to
singular character of the $\bf M$--matrix, which is irrelevant to
the thermodynamical properties of the physical system. Thus, we
can write the partition function as
\begin{eqnarray}\label{eq34x}
Z=N'N(\beta)\! \int_{_{\hspace{-0.5cm}\mbox{\footnotesize
periodic}}} \hspace{-0.4cm}{\cal D}\psi \,{\cal D}\overline\psi\;
\;\exp\left[\int_0^\beta \!\!\! d\tau\!\!\int\!d^3{\bf x}
\;\overline\psi({\bf D}+ \mu\beta^0)\psi \,\right].
\end{eqnarray}

The fields may be expanded as a Fourier series \cite{bernard1}
\begin{eqnarray} \label{eq35x}
\psi(\tau,{\bf x})&=&\psi_{zero}+\frac{1}{\beta}\, \sum_{n}
\!\int\!\!\frac{d^3\bf p}{(2\pi)^3}\;e^{i({\bf p}\cdotp{\bf x} +
\omega_n\tau)}\psi_n({\bf p})\nonumber\\ \\
\overline\psi(\tau,{\bf x})&=&\overline\psi_{zero} +
\frac{1}{\beta}\, \sum_{n} \!\int\!\!\frac{d^3\bf
p}{(2\pi)^3}\;e^{i({\bf p}\cdotp{\bf x} + \omega_n\tau)}
\overline\psi_n({\bf p})\nonumber\;,
\end{eqnarray}
where $\psi_{zero}$ and $\overline\psi_{zero}$ are independent of
$(\tau,{\bf x})$ and carry the full infrared character of the DKP
field; that is, $\psi_{n=0}({\bf p}=0)=0$ and $\overline\psi_{n=0}
({\bf p} =0)=0$. And due to bosonic character of DKP field, we
impose the constraint of periodicity such  $\psi(\tau=0,{\bf
x})=\psi(\tau=\beta,{\bf x})$ for all $\bf x$, thus the frequency
will be $\omega_n=2\pi n/\beta$, $n=\pm 1,\, \pm 2,\,...$.

Substituting (\ref{eq35x}) into (\ref{eq34x}), we find
\begin{eqnarray}
\int_0^\beta\! \!\!\! d\tau\!\!\int \!d^3{\bf x} \;
\overline\psi({\bf D}+ \mu\beta^0)\psi=
\beta\,V\overline\psi_{zero}(-m+\mu\beta^0) \psi_{zero} \;+\;S'\;,
\end{eqnarray}
where the free infrared action is
\begin{eqnarray}\label{eq37x}
S'\;=\;-\frac{1}{\beta}\, \sum_{n} \!\int\!\!\frac{d^3\bf
p}{(2\pi)^3}\;   \overline\psi_{-n}(-\mathbf{p}) \,\!\left[
i\,\beta^0(\omega_n+i\,\mu)+\beta^k p_k +m\right]
\psi_{n}(\mathbf{p})\;.
\end{eqnarray}

The partition function given in (\ref{eq34x}), can be rewritten
using the equation (\ref{eq37x}) as
\begin{eqnarray}\label{eq38x}
Z=N' N(\beta) \,Z_{zero}\,Z'\;,
\end{eqnarray}
where $Z_{zero}$ is the zero mode contribution
\begin{eqnarray} \label{eq39x}
Z_{zero}=\exp\left[\frac{}{}\beta\,V\overline\psi_{zero} (-m+
\mu\beta^0) \psi_{zero}\right]\,,
\end{eqnarray}
and the $Z'$ contribution is
\begin{eqnarray}\label{eq40x}
Z'=\prod_n \prod_{\bf p}\,\left[\det\left(\frac{}{}i\,\beta^0
(\omega_n+i\,\mu)+\beta^k p_k +m\,\right)\right]^{-1}\,.
\end{eqnarray}
\subsection{The spin 0 sector}

It is worth to notice that until this moment we have not used an
explicit representation of the DKP algebra. In this point we
specialize in the spin 0 sector of DKP theory, thus we calculate
every contribution in (\ref{eq38x}). We first compute the
(\ref{eq40x}) $Z'$ partition function, to make it we use an
explicit $5\times 5$ representation (\ref{eqa1}) (see Appendix) of
the DKP algebra, thus we obtain
\begin{eqnarray}\label{eq41x}
\ln Z'=-V\sum_n \!\int\!\!\frac{d^3\bf p}{(2\pi)^3}\;
\ln\left\{m^3\left[\left(\omega_n+i\mu
\right)^2+\omega^2\right]\right\}\,,
\end{eqnarray}
where $\omega=\sqrt{{\bf p}^2+m^2}$, and ignoring the
$\beta$--independent constant, we can rewrite (\ref{eq41x}) to
show explicitly the existence of particles and anti-particles,
thus
\begin{eqnarray}\label{eq42x}
\ln Z'& \!\!\!\!=&\!\!\!\!2V\ln(\beta)\sum_n\!\int\!\!\frac{d^3\bf
p}{(2\pi)^3}\; -\frac{V}{2}\sum_n\!\int\!\!\frac{d^3\bf
p}{(2\pi)^3}\;\ln\left\{ \beta^2\left[\omega^2_n + \left(\omega-\mu
\right)^2\right]\right\}+\nonumber\\& &
-\frac{V}{2}\sum_n\!\int\!\!\frac{d^3\bf p}{(2\pi)^3}
\;\ln\left\{\beta^2 \left[\omega^2_n+\left(\omega+\mu
\right)^2\right]\right\}\,.
\end{eqnarray}

To calculate the sum on $n$ in the last two terms we use the
standard results founded in the literature
\cite{kapusta1,bernard1}, thus
\begin{eqnarray} \label{eq43x}
\ln Z'&\!\!\!\!=&\!\!\!\!-V\int\!\!\frac{d^3\bf p}{(2\pi)^3}
\left[\frac{}{} \! \beta\omega+\ln\left(1-e^{-\beta(\omega-\mu)}
\right)+\ln\left(1-e^{-\beta(\omega+\mu)}\right)\right] +
\nonumber\\
& &  +2V\ln(\beta)\sum_n\!\int\!\!\frac{d^3\bf p}{(2\pi)^3}\;,
\end{eqnarray}
where the last term is an infinity $\beta$--dependent quantity
which has to be cancelled by the $N(\beta)$ constant, thus we set
it to be
\begin{eqnarray} \label{eq43x-1}
\ln N(\beta)=-2V\ln(\beta)\sum_n\!\int\!\!\frac{d^3\bf
p}{(2\pi)^3}
\end{eqnarray}
in total agreement with the literature when is studied the complex
or charged scalar field \cite{kapusta1,weldon,lebellac1}.

\subsubsection*{The zero mode contribution} 

Now, we show the existence of the zero mode sector for the spin 0
DKP field. The zero mode equation is obtained from the action given
in
(\ref{eq34x}),
\begin{eqnarray} \label{eq44x}
\left(\frac{}{}\!\mathbf{D}+\mu\beta^0\right)\psi=0
\end{eqnarray}
by setting the fields $\psi$ to be independent of $(\tau,{\bf
x})$, thus
\begin{eqnarray}\label{eq45x}
\left(\frac{}{}\!-m+\mu\beta^0\right)\psi_{zero}=0
\end{eqnarray}
then the determinant of the matrix $\left( -m+ \mu\beta^0\right)$
must be set to zero for getting non--trivial solutions,
\begin{eqnarray} \label{eq46x}
\det \left( -m+ \mu\beta^0\right)=-m^3 (m^2-\mu^2)=0\,.
\end{eqnarray}
It is easy to see that we get non-trivial solutions when the
chemical potential $\mu$ reaches the values $\pm m$. The field
$\psi_{zero}$ is a five complex component column matrix,
\begin{eqnarray}\label{eq47x}
\psi_{zero}=\left(a,b,c^1,c^2,c^3\right)^T\,.
\end{eqnarray}
Solving the matrix equation above, we find
\begin{eqnarray}\label{eq48x}
c^k = 0 \quad,\quad a=\frac{\mu b}{m} \quad \mbox{or} \quad
a=\frac{m b}{\mu}\;.
\end{eqnarray}
Then, using the above relations we find that zero mode
contribution of the partition function is for $\mu \neq \pm m$
\begin{eqnarray} \label{eq49x}
Z_{zero}=-\beta\,V (m^2-\mu^2)\xi^2
\end{eqnarray}
with $\xi$ being an arbitrary real parameter.

Finally, the full partition function for the spin 0 sector of DKP
field is
\begin{eqnarray} \label{eq50x}
\ln Z&\!\!\!\!=&\!\!\!\!\!- \beta\,V (m^2\!-\mu^2)\xi^2\!
-V\!\!\int\!\!\frac{d^3\bf p}{(2\pi)^3}\! \left[\frac{}{} \!
\beta\omega+\ln\!\left(1- e^{-\beta(\omega-\mu)}\right)\!+
\ln\!\left(1-e^{-\beta(\omega+\mu)} \right)\right] \quad
\end{eqnarray}
reproducing exactly the partition function of charged scalar field
\cite{kapusta1} when we take the contribution of the zero mode into
account.  The momentum integral is convergent only if $|\mu|\leq m$.
The parameter $\xi$ that appears in the final  expression for the
partition function is not determined {\it a priori}, it should be
treated as an variational parameter which is related to the charge
carried by the condensed particles. Then, at fixed $\beta$ and
$\mu$, $\ln Z$ is an extremum with respect to variations of such
a free parameter,
\begin{eqnarray}\label{eq50-x1}
\frac{\partial\,\ln Z}{\partial \xi}=-2\beta V(m^2-\mu^2) \xi =0
\end{eqnarray}
which implies that $\xi=0$, unless $|\mu|=m$ and in this case
$\xi$ is undetermined by this variational condition. Thus, when
$|\mu|<m$ the contribution of the zero mode to the partition
function is cancelled.

\subsubsection{The Bose-Einstein condensation}

By completeness, we briefly summarize \cite{weldon} some
characteristic of Bose--Einstein condensation in the case of a
relativistic Bose gas. The question which we make is this: What is
the requirement for BEC to take place at relativistic temperatures
(i.e., $T\gg m$) and what is the nature of the phase transition.

For $|\mu|<m$, the $\xi$ parameter is set to zero in (\ref{eq50x}),
and the charge density is given by
\begin{eqnarray}\label{eq51-1x}
\rho=\frac{1}{\beta V}\;\left(\frac{\partial\,\ln Z}{\partial \mu}
\right)=\int\!\!\frac{d^3\bf p}{(2\pi)^3} \left(\frac{1}{
e^{\beta(\omega-\mu)}-1}-\frac{1}{e^{\beta (\omega+\mu)}-1}\right),
\end{eqnarray}
as the chemical potential satisfies $|\mu|<m$, thus, the charge
densities of particles and anti--particles are non--negative.
Note that (\ref{eq51-1x}) is really an implicit formula for $\mu$
as a function of $\rho$ and $T$. For $T$ above the some critical
temperature $T_{_C}$, one can always find a $\mu$ such that
(\ref{eq51-1x}) holds. If the density $\rho$ is held fixed and the
temperature is lowered, $\mu$ will increase until the point
$|\mu|=m$ is reached, thus, in the region $T\geq T_{_C}\gg m$ we
obtain \cite{weldon}
\begin{eqnarray}\label{eq51-1x1}
|\rho|\approx\frac{1}{3}\;m T^2
\end{eqnarray}

When $|\mu|=m$ and the temperature is lowered even further such
$T<T_{_C}$, the charge density is written as
\begin{eqnarray}\label{eq51-2x}
\rho= \frac{1}{\beta V}\;\left(\frac{\partial\,\ln Z}{\partial \mu}
\right)_{\mu=m} =\rho_0 + \rho^*(\beta,\mu=m)\,,
\end{eqnarray}
where $\rho_0=2m\xi^2$ is a charge contribution from the condensate
(the zero--momentum mode) and the $\rho^*(\beta,\mu=m)$ is the
thermal particle excitations (finite--momentum modes) which is
given by (\ref{eq51-1x}) with $|\mu|=m$.

The critical temperature $T_{_C}$, in which the Bose--Einstein
condensation occurs, corresponds when  $|\mu|=m$ is reached and
determined implicitly by the equation
\begin{eqnarray}\label{eq51-1x2}
\rho=\rho^*(\beta_C,\mu=m)\,,
\end{eqnarray}
which implies that
\begin{eqnarray}\label{eq51-3}
T_{_C}=\left(\frac{3|\rho|}{m}\right)^{1/2} \;.
\end{eqnarray}
At temperatures $T<T_{_C}$, (\ref{eq51-2x}) is an equation for
charge density $\rho-\rho_0$ of the $\bf p \neq 0$ states,
\begin{eqnarray}\label{eq51-4}
\rho-\rho_0=\frac{1}{3}\;m T^2\,,
\end{eqnarray}
so that the charge density in the ground state ($\bf p = 0)$ is
\begin{eqnarray}\label{eq51-5}
\rho_0=\rho\left(1-\left[\frac{T}{T_{_C}}\right]^2\right)\,,
\end{eqnarray}
as the temperature is lowered the fraction of charge in $\bf p = 0$
state increase to unity. We note that (\ref{eq51-3}) leads to the
important result: the necessary condition for that any ideal Bose
gas of mass $m$ Bose--Einstein condense at a relativistic
temperature (i. e., $T_{_C}\gg m$), is that $\rho \gg m^3$.
Additional literature about Bose--Einstein condensation can be found
in \cite{huang,kapusta2,lebellac1,bec1x}

\subsection{The spin 1 sector}

Now we study spin 1 sector of DKP theory, thus we calculate every
contribution in (\ref{eq38x}). We first compute the (\ref{eq40x})
$Z'$ partition function. For this purpose, we consider an explicit
$10\times 10$ representation (\ref{eqa2}) (see Appendix) of the DKP
algebra, thus we get the following contribution
\begin{eqnarray}\label{eq53x}
\ln Z'=-V\sum_n \!\int\!\!\frac{d^3\bf p}{(2\pi)^3}\;
\ln\left\{m^4\left[\left(\omega_n+i\mu\right)^2+\omega^2\right]^3
\right\}
\end{eqnarray}
and more one time by ignoring the $\beta$--independent constant,
we can rewrite (\ref{eq53x}) to show explicitly the existence of
particles and anti-particles, thus
\begin{eqnarray}\label{eq54x}
\ln Z'&\!\!\!\!=&\!\!\!\!6V\ln(\beta)\sum_n\!\int\!\!\frac{d^3\bf
p}{(2\pi)^3}\; -\frac{3V}{2}\sum_n\!\int\!\!\frac{d^3\bf
p}{(2\pi)^3}\;\ln\left\{ \beta^2\left[\omega^2_n +
\left(\omega-\mu\right)^2\right]\right\}+\nonumber\\& &
-\frac{3V}{2}\sum_n\!\int\!\!\frac{d^3\bf p}{(2\pi)^3}
\;\ln\left\{\beta^2 \left[\omega^2_n+\left(\omega+\mu
\right)^2\right]\right\}.
\end{eqnarray}
Here, we can see the six degree of freedom corresponding to two
complex  massive vector fields or Proca's complex fields. Carrying
out the sum in both last terms and after some manipulations
\cite{kapusta1,bernard1}, we find
\begin{eqnarray} \label{eq55x}
\ln Z'&\!\!\!\!=&\!\!\!\!-3V\int\!\!\frac{d^3\bf p}{(2\pi)^3}
\left[\frac{}{} \! \beta\omega+\ln\left(1-e^{-\beta(\omega-\mu)}
\right)+\ln\left(1-e^{-\beta(\omega+\mu)}\right)\right] + \nonumber
\\ & &  +6V\ln(\beta)\sum_n\! \int\!\!\frac{d^3\bf p}{(2\pi)^3}\;,
\end{eqnarray}
the last term is an infinity quantity $\beta$--dependent, and, as
in the spin 0 sector, it will be cancelled by the $N(\beta)$
contribution, thus with all security we set it to be
\begin{eqnarray} \label{eq55x-1}
\ln N(\beta)=-6V\ln(\beta)\sum_n\!\int\!\!\frac{d^3\bf
p}{(2\pi)^3}\;.
\end{eqnarray}

\subsubsection*{The zero mode contribution} 

In this case, we also show the existence of the zero mode sector
for the spin 1 DKP  field. The zero mode equation is obtained from
the action given in
(\ref{eq34x}),
\begin{eqnarray} \label{eq56x}
\left(\frac{}{}\!\mathbf{D}+\mu\beta^0\right)\psi=0
\end{eqnarray}
by setting the fields $\psi$ to be independent of $(\tau,{\bf
x})$, thus
\begin{eqnarray}\label{eq57x}
\left(\frac{}{}\!-m+\mu\beta^0\right)\psi_{zero}=0
\end{eqnarray}
then o determinant of the matrix $\left( -m+ \mu\beta^0\right)$
must be set to zero for getting non--trivial solutions,
\begin{eqnarray} \label{eq58x}
\det \left( -m+ \mu\beta^0\right)=m^4 (m^2-\mu^2)^3=0\;.
\end{eqnarray}
It is easy to see that we get non-trivial solutions when the
chemical potential $\mu$ reaches the values $\pm m$, and in this
representation the field $\psi_{zero}$ is a ten complex component
column matrix,
\begin{eqnarray}\label{eq59x}
\psi_{zero}=\left(\psi^\tau ,\psi^1,\psi^2,\psi^3,\psi^4,\psi^5,
\psi^6,\psi^7,\psi^8,\psi^9\right)^T\,.
\end{eqnarray}
By solving the above matrix equation, we find
\begin{eqnarray}\label{eq60x}
\psi^\tau=\psi^4&\!\!\!=&\!\!\!\psi^5=\psi^6 = 0 \nonumber\\ \\
\psi^1=\frac{\mu \psi^7}{m} \quad,\quad \psi^2&\!\!\!=&\!\!\!
\frac{\mu \psi^8}{m} \quad,\quad \psi^3=\frac{\mu \psi^9}{m}
\nonumber\;.
\end{eqnarray}
By using the above relations we find that zero mode contribution
of the partition function is for $\mu \neq \pm m$
\begin{eqnarray} \label{eq61x}
Z_{zero}=-\beta\,V
(m^2-\mu^2)\left(\frac{}{}\!\xi^2_1+\xi^2_2+\xi^2_3 \right)
\end{eqnarray}
with $\xi_k$ being an arbitrary real parameters related to the
spatial components to the zero mode DKP field, i.e, $\psi^1,\,
\psi^2,\,\psi^3$.

Finally, the full partition function for the spin 1 sector of DKP
field is
\begin{eqnarray} \label{eq62x}
\ln Z&\!\!\!=&\!\!\!-\beta\,V (m^2 -\mu^2)
\left(\xi^2_1+\xi^2_2+\xi^2_3\right)+\\ & &
 -3V\!\!\int\!\!\frac{d^3\bf p}{(2\pi)^3}\! \left[\frac{}{} \!
\beta\omega+\ln\!\left(1- e^{-\beta(\omega-\mu)}\right)\!+
\ln\!\left(1-e^{-\beta(\omega+\mu)} \right)\right]\nonumber .
\end{eqnarray}

The $\xi_k$ parameters are related to the degree of freedom of
the spin 1 DKP field, i.e, three degrees of freedom of the massive
charged vector field. The zero mode sector is again absolutely
necessary to reproduce the BEC as the spin zero sector; however,
the critical temperature is the same for every degree of freedom.


\section{Conclusions and remarks}

The free massive DKP theory at Finite Temperature (FT) is
equivalent to both complex massive scalar field and complex vector
field theories at FT. It reproduces the relativistic Bose--Einstein
condensation in both sectors, where we show in a clean and elegant
way the zero mode existence and its contribution to BEC.

The perspectives to follow  are to study the DKP field coupled to
the quantized electromagnetic field and the implementation of the
renormalization process at zero temperature which  will allow to
extend the analysis at Finite temperature. And,  Bose--Einstein
condensation  in curved space--times \cite{bec-cur} using the DKP
theory minimally coupled. Advances in this directions will be
reported elsewhere.


\subsection*{Acknowledgements}

We thank to Profs. L. Tomio and Victo S. Filho for additional
references related to  the Bose--Einstein condensation. This work
was supported by FAPESP/Brazil (R.C., full support grant 01/12611-7;
V.Ya.F., grant 01/12585-6; B.M.P., grant 02/00222-9, J. V. full
support grant 00/03812-6), RFFI/ Russia (V.Ya.F., grant 02-02-16946),
LSS-1578.2003.2 (V.Ya.F.),
and CNPq/Brazil (B.M.P.).

\newpage
\appendix
\section*{Appendix}

\subsection*{Spin 0 and Spin 1 representation}

We use the following representation for the spin 0 sector of DKP
algebra
\begin{eqnarray} \label{eqa1}
\hspace{-5.9cm}\scalebox{0.75}{\includegraphics[86,552][300,650]
{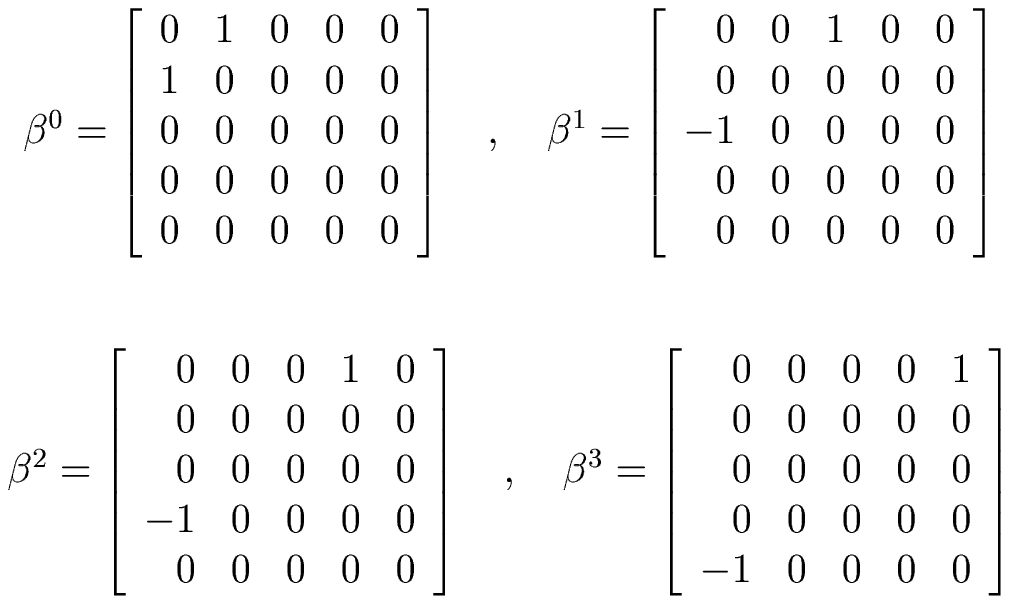}}
\end{eqnarray}
\vskip 2.4cm \hspace{-0.5cm}and for the spin 1 sector of DKP algebra
\begin{eqnarray} \label{eqa2}
\hspace{-7.5cm}\scalebox{0.75}{\includegraphics[100,512][300,660]
{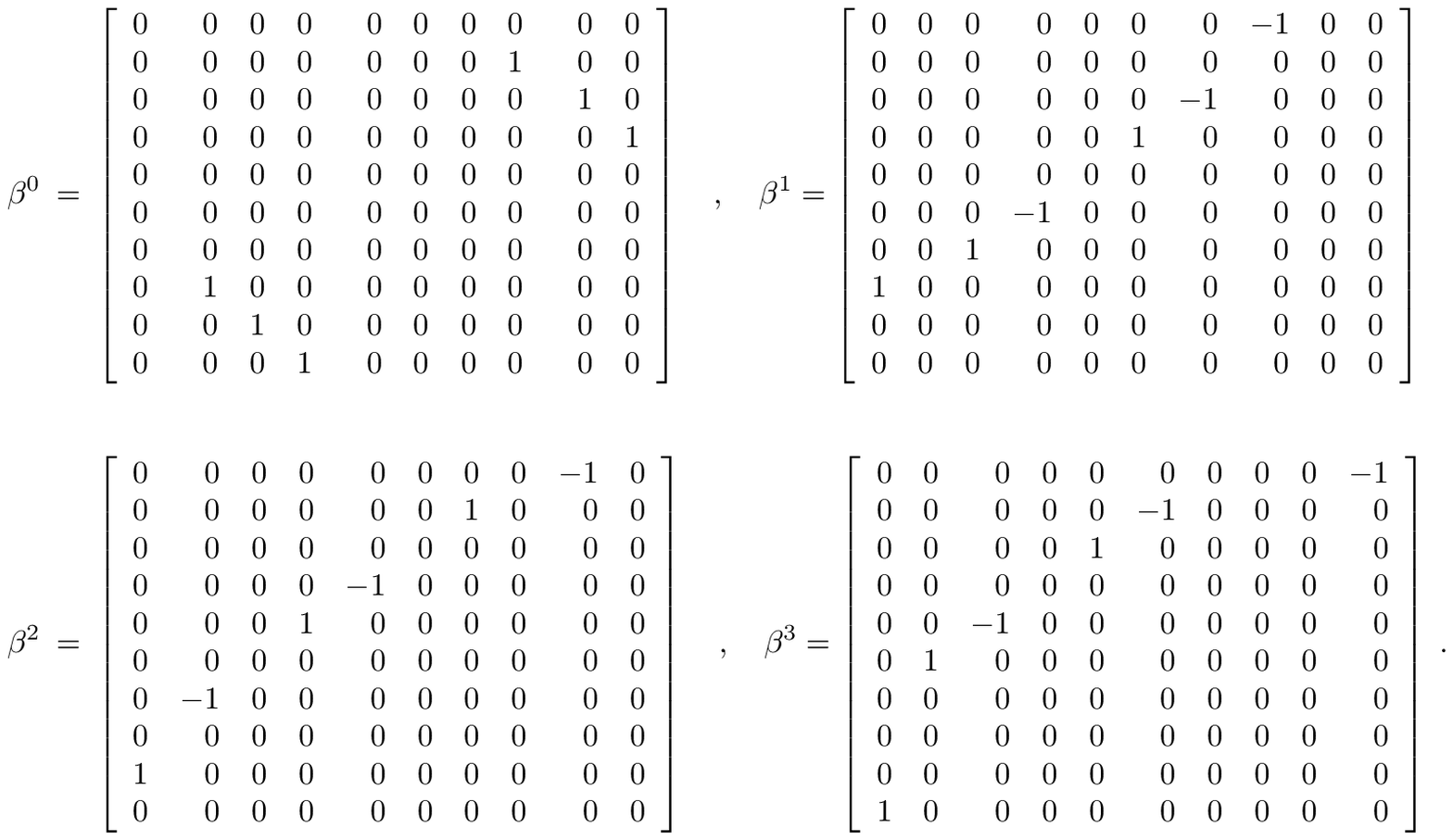}}
\end{eqnarray}

\end{document}